\documentclass[11pt,a4paper,english]{article}
\usepackage{jstyle}
\usepackage{babel,csquotes}
\usepackage{amssymb,amsmath}
\usepackage{tensor}
\usepackage{tikz}
\usepackage{hyperref}

\author[a]{Euihun JOUNG,}
\author[a]{Wenliang LI}
\author[b,c]{and Massimo TARONNA}

\affiliation[a]{AstroParticule et Cosmologie\footnote{Unit\'e Mixte de Recherche 7164 du CNRS}\\ 
10 rue Alice Domon et L\'eonie Duquet, 75205 Paris Cedex 13, France}

\affiliation[b]{Max-Planck-Institut f\"ur Gravitationsphysik
(Albert-Einstein-Institut)\\
Am M\"uhlenberg 1, 14476 Golm, Germany}

\affiliation[c]{Kavli Institute for Theoretical Physics China, CAS\\
Beijing 100190, China}

\emailAdd{joung@apc.univ-paris7.fr}
\emailAdd{wenliang.li@apc.univ-paris7.fr}
\emailAdd{massimo.taronna@aei.mpg.de}

\title{\centering No-Go Theorems for Unitary and Interacting Partially Massless Spin-Two Fields
}

\abstract{We examine the generic theory of a partially massless (PM) spin-two field interacting with gravity in four dimensions from a bottom-up perspective. By analyzing the most general form of the Lagrangian, we first show that if such a theory exists, its de Sitter background must admit either $\mathfrak{so}(1,5)$ or 
$\mathfrak{so}(2,4)$ global symmetry depending on the relative sign of the kinetic terms: the former for a positive sign the latter for a negative sign. Further analysis reveals that the coupling constant of the PM cubic self-interaction must be fixed with a purely imaginary number in the case of a positive sign. We conclude that there cannot exist a unitary theory of a PM spin-two field coupled to Einstein gravity with a perturbatively local Lagrangian. In the case of a negative sign we recover conformal gravity. As a special case of our analysis, it is shown that the PM limit of massive gravity also lacks the PM gauge symmetry.
}

\begin{document}

\pagestyle{empty}
\hfill AEI-2014-027
\vskip 0.1\textheight

\maketitle

\section{Introduction}

In de Sitter space (dS) space, unitary spin-two modes have a mass gap, 
as opposed to those in the flat space or anti-de Sitter space. The lightest massive spin-two modes do not correspond to the massless graviton but to
a special massive field called \emph{partially massless} spin-two \cite{Deser:1983mm,Deser:2001us}. This lower bound is also known as \emph{Higuchi bound} \cite{Higuchi:1986py}.
The partially-massless spin-two (PM) field has one less degree of freedom (DoF) than a generic massive spin-two field due to the decoupling of the scalar mode: 
for example in four dimensions, it has four DoFs instead of the five of the usual massive field.

The PM field is gaining renewed interest in the context of the massive gravity theory of
\cite{deRham:2010ik,deRham:2010kj,Hassan:2011vm} and the bimetric gravity theory of \cite{Hassan:2011ea,Hassan:2011zd}.
With a suitable choice of parameters,
these theories can be linearized around dS space
and describe the propagation of massive spin-two modes.
One of the natural questions is the following: when the mass is tuned to that of PM \footnote{The possibility of having partially massless fields in the context of massive gravity has been first discussed in 3D massive gravity \cite{Bergshoeff:2009aq}.}, can the resulting theory consistently describe the dynamics of an interacting PM field? 
In other words, does the scalar DoF decouple from the theory in the PM limit?
In the free theory of the PM field $\varphi_{\m\n}$\,, the decoupling of the scalar DoF is due to
the emergence of a gauge symmetry of the form,
\be
	\delta\,\varphi_{\mu\nu}=\left(\bar\nabla_{\m}\bar\nabla_{\n}+\frac{\L}3\,\bar g_{\m\n}\right)\a\,,
	\label{PM sym lin}
\ee
where $\bar g_{\m\n}$ and $\bar\nabla_{\m}$ are the metric and covariant derivative of dS space
with cosmological constant $\L$\,.
If the PM limit of massive or bimetric gravity is consistent,
then they should also admit a PM gauge symmetry
which extends the free one \eqref{PM sym lin} to the interacting level.
While the emergence of such a gauge symmetry has not yet been reported,
there have been many discussions on the possible (in-)consistencies of this limit: 
see \cite{deRham:2012kf,Hassan:2012gz,Hassan:2012rq,Hassan:2013pca} for positive and \cite{Deser:2012qx,Deser:2013uy,deRham:2013wv,Deser:2013gpa,Fasiello:2013woa} for negative results.
One of the aims of the present work is to provide a definite answer to this question.

Another playground for the PM field is conformal gravity (CG),
which has six propagating DoFs \cite{Fradkin:1981iu,Lee:1982cp,Riegert:1984hf}. 
Two DoFs correspond to the usual graviton while the additional four DoFs 
organize themselves into a PM representation around dS space (see \emph{e.g.}
\cite{Maldacena:2011mk,Deser:2012qg}).
In order to see this point, it is convenient to recast the action into 
\be
	S_{\rm\sst CG}
	=\int d^{4}x \sqrt{-g}\left[
	-\frac\L6\,(R-2\,\L)+\frac6\L\,\mathscr L_{\sst\rm PM}(\varphi,\nabla \varphi,g,R) \right],
	\label{CG}
\ee
by introducing an \emph{auxiliary} field $\varphi_{\mu\nu}$ which can then be interpreted as a PM field.
Here, $\mathscr L_{\sst\rm PM}$ 
is a Lagrangian whose quadratic part coincides with that of the free PM field, while the higher power parts involve interactions
of $\varphi_{\mu\nu}$ --- see \cite{Deser:2012qg} for more details. 

CG is non-unitary because of the wrong relative sign
between the Einstein-Hilbert term and $\mathscr L_{\sst\rm PM}$\,.
Nevertheless, as far as the number of DoFs is concerned, CG provides
a consistent theory of PM plus gravity.
One can see this from the presence of the PM gauge symmetry in CG,
which is nothing but a disguised version of Weyl symmetry.
Since the relative negative sign makes the theory non-unitary,
one may think of naively flipping this sign.
The redefinition $\varphi_{\mu\nu}\to \pm i\,\varphi_{\mu\nu}$ might do the job, 
but it would introduce an imaginary part in $\mathscr L_{\sst\rm PM}$\,, due to the presence of 
odd-power interactions involving PM fields.
Hence, there is no obvious simple way to obtain a unitary theory of PM plus gravity 
out of CG. 

In the search of a consistent theory of interacting PM field, 
it is instructive to try to build it starting from the free theory order by order in powers of the PM field. 
About two-derivative cubic interactions,
the task has been carried out in \cite{Zinoviev:2006im}\footnote{See also 
 \cite{Zinoviev:2013hac,Zinoviev:2014zka}
for the other related results of the same author.} while the cubic interactions with general number of derivatives
have been analyzed in a wider context in \cite{Joung:2012rv,Joung:2012hz} for the fields of arbitrary spins and arbitrary masses in generic dimensions. Let us briefly summarize the results: 
\begin{itemize}
\item {\bf Two-derivative couplings:\quad}
The cubic self-interaction of PM field is unique and exists only in four dimensions.
It is associated with Abelian conserved charges.
Moreover, it coincides with the coupling that can be extracted from the CG action \eqref{CG} \cite{Deser:2012qg}.
The PM--PM--graviton interaction, which corresponds to the gravitational minimal coupling,
is consistent provided that the graviton also transform under PM transformation.
It is associated with non-Abelian conserved charges.
\item {\bf Higher-derivative couplings:\quad}
Any of the cubic self-interactions of PM field with more than two derivatives
is not associated with a conserved charge.
\end{itemize}

In the present paper, we prove that there cannot exist a unitary interacting theory for PM plus gravity.
Our proof is based on the following two consequences of the gauge invariance condition:
\begin{itemize}
\item {\bf Global symmetry:}\quad Conserved charges must form a Lie algebra.
\item {\bf Admissibility:}\quad The linearized theory must carry a unitary representation of the global symmetry algebra.  
\end{itemize}
We first show, by demanding PM field to be gravitationally interacting, that the first condition is automatically satisfied,
and that the corresponding global symmetry turns out to be $\mathfrak{so}(1,5)$ containing the dS isometry algebra 
$\mathfrak{so}(1,4)$.
Since the cubic PM self-interaction does not give rise to any non-Abelian charge,
its coupling constant, say $\lambda$\,, does not enter in the structure constants of $\mathfrak{so}(1,5)$\,.
Moving to the second requirement, \emph{admissibility condition}, we demonstrate that the linearized field can carry a representation of 
$\mathfrak{so}(1,5)$ only when the PM self-interaction coupling constant $\l$ satisfies
\be
	\lambda^{2}+8\,\pi\,G_{N}=0\,,
\ee
where $G_{N}$ is the Newton's gravitational constant. This shows that the gauge invariance of 
PM plus gravity action requires the PM self interaction to have an imaginary coupling constant
$\l=\pm i\,\sqrt{8\,\pi\,G_{N}}$, which manifestly violates the unitarity.
In the case of PM theory without gravity, which can be achieved by taking \mt{G_{N}\to 0} limit,
we get $\lambda=0$ so the theory cannot have a cubic interaction. 
In particular, this implies that the PM limit of massive/bimetric gravity cannot lead to a gauge invariant 
Lagrangian theory, and consequently 
it should suffer from a kind of Boulware-Deser ghost problem 
\cite{Boulware:1973my}: the scalar DoF does not decouple from the theory.
Moreover, one can notice that this choice of imaginary $\l$ coincides with the redefinition
$\varphi_{\m\n} \to \pm i\,\varphi_{\m\n}$ of CG:
in such a case, the global symmetry $\mathfrak{so}(1,5)$ is replaced
by the conformal algebra $\mathfrak{so}(2,4)$.
Therefore, we conclude that the only interacting theory of PM field and gravity is CG, irrespectively of
the unitarity issue.

The organization of the present paper is as follows. 
In \autoref{sec: Grav}, we describe the bottom-up construction of the PM plus gravity action together with the associated deformations of gauge symmetries. 
In \autoref{Sec: symmetries}, we examine the induced symmetries of this theory and the corresponding representation arriving to our main results.
Finally, \autoref{sec: Conclusions} contains our conclusion
while some additional materials can be found in the appendices.

\section{Gravitationally interacting PM field}
\label{sec: Grav}

In search for a theory of PM plus gravity, 
one may begin with the most general form of the action $S$\,. It has two parts:
\be
	S=S_{\rm\sst EH}+S_{\rm\sst PM}\,,
	\label{Action}
\ee
where the gravity sector $S_{\rm\sst EH}$ is given by Einstein-Hilbert term:
\be
	S_{\rm\sst EH}[g]=\frac{1}{2\,\k}\int d^{4}x\,\sqrt{-g}\ (R-2\,\L)\,,
\ee
with $\k=8\,\pi\,G_{N}$\,, while the PM part $S_{\sst\rm PM}$ is not fixed for the moment
except that it is given through a quasi-local\footnote{By \emph{quasi-local} Lagrangian,
we mean that there exists an expansion parameter such that every truncation of the Lagragian to a finite power of this parameter contains finitely many derivatives. The number
of derivatives of the full Lagrangian may be unbounded.} Lagrangian $\mathscr L_{\rm\sst PM}$
of manifestly diffeomorphism-invariant form:
\be
	S_{\rm\sst PM}[\varphi,g]
	=\int d^{4}x\,\sqrt{-g}\ \mathscr{L}_{\rm\sst PM}(\varphi,\nabla \varphi,g,R,\ldots)\,,
\ee
where, $\ldots$ means that there may be higher derivatives of $\varphi_{\m\n}$ or curvature $R_{\m\n\r\s}$. 
Let us emphasize that this ansatz also covers the bimetric gravity of \cite{Hassan:2011ea,Hassan:2011zd}:
see \autoref{sec: bimetric} for more details.

Besides the diffeomorphism symmetries, we also require the action to be invariant 
under PM gauge symmetries:
\be
	\delta_{\alpha}\, S=0\,,
	\label{PM inv}
\ee 
where $\delta_{\a}$ is the nonlinearly deformed PM transformation which we 
aim to determine together with $\mathscr L_{\sst\rm PM}$\,.
For further analysis of this gauge invariance condition, it is convenient to expand the action and the PM gauge transformations in powers of the PM field $\varphi_{\mu\nu}$ as
\be
	S_{\rm\sst EH}=S^{\sst (0)}\,,
	\qquad
	S_{\rm\sst PM}=S^{\sst (2)}+S^{\sst (3)}+\cdots\,,\qquad
	\delta_{\alpha}=\delta^{\sst (0)}_{\alpha}+\delta^{\sst (1)}_{\alpha}+\cdots\,,
	\label{Expan}
\ee
where
\be
	\delta^{\sst (n)}_{\alpha}=\int d^{4}x\sqrt{-g}\left[\left(\delta_{\a}\varphi_{\m\n}\right)^{\sst (n)}
	\,\frac{\d\ }{\d \varphi_{\m\n}}
	+\left(\delta_{\a}g_{\m\n}\right)^{\sst (n-1)}\frac{\d\ }{\d g_{\m\n}}\right],
\ee
while the superscript $(n)$ means that the corresponding term involves the $n$th power of $\varphi_{\m\n}$\,.
Then, the PM gauge invariance condition \eqref{PM inv}  provides an infinite set of equations:
\ba
	&&\delta^{\sst (1)}_{\alpha} S^{\sst (0)}=0\,,\label{cond1}\\ 
	&&\delta^{\sst (0)}_{\alpha} S^{\sst (2)}
	+\delta^{\sst (2)}_{\a} S^{\sst (0)}=0\,,\label{cond2}\\
	&&\delta^{\sst (0)}_{\alpha} S^{\sst (3)}
	+\delta^{\sst (1)}_{\alpha} S^{\sst (2)}
	+\delta^{\sst (3)}_{\a} S^{\sst (0)}=0\,,\label{cond3}\\
	&& \qquad \cdots\,.\nonumber
\ea
The first condition \eqref{cond1} simply tells us that $(\delta_{\a} g_{\m\n})^{\sst (0)}=0$ --- the metric does not transform under PM 
at the lowest order ---
whereas the other conditions constrain possible forms of $\mathscr L_{\sst\rm PM}$
and $\delta_{\a}$\,.
The advantage of the expansion \eqref{Expan} is that we can  attack the gauge invariance conditions
one by one from the lowest level.
In the following, we shall analyze the second  and third  conditions (\ref{cond2}\,,\,\ref{cond3}).

\subsection{Quadratic part}

The quadratic part of the gauge invariance condition \eqref{cond2} reads
\be
	\int d^{4}x\sqrt{-g}
	\left( \left(\delta_{\a}\,\varphi_{\mu\nu}\right)^{\sst (0)}
	\left[\frac{\delta S^{\sst (2)}}{\delta \varphi_{\mu\nu}}\right]
	+ \left(\delta_{\a} g_{\mu\nu}\right)^{\sst (1)} G^{\mu\nu}_{\L} \right)=0\,,
	 \label{Cond1}
\ee
where $G^{\m\n}_{\L}\equiv R^{\mu\nu}-g^{\mu\nu}\,R/2+\L\,g^{\mu\nu}$ is the cosmological 
Einstein tensor.
The lowest-order PM gauge transformation is given by the covariantization of the free PM transformation \eqref{PM sym lin} around the dS background:
\be
	\left(\delta_{\a}\varphi_{\mu\nu}\right)^{\sst (0)}=\left(\nabla_{\m}\nabla_{\n}+\frac{\L}3\,g_{\m\n}\right)\a\,,
	\label{PM zero}
\ee
up to trivial transformations proportional to $G^{\m\n}_{\L}$\,.
Eq.~\eqref{cond2} can be solved by properly covariantizing the free action of the PM field around dS background. The solution for $S^{\sst (2)}$ reads
\ba
	S^{\sst (2)}
	\eq  \s \int d^{4}x\sqrt{-g}\,\bigg[-\frac 1 2 \,\nabla_\alpha \varphi_{\mu\nu} \, \nabla^\alpha \varphi^{\mu\nu} 
	+  \nabla_\alpha \varphi_{\mu\nu} \, \nabla^\nu \varphi^{\mu\alpha}
	-  \nabla_\mu\varphi \, \nabla_\nu \varphi^{\mu\nu} 
	+ \frac 1 2 \, \nabla_\mu \varphi \, \nabla^\mu \varphi \nn
	&&\qquad+\,\L\left(\varphi_{\mu\nu}\,\varphi^{\mu\nu} - \frac 1 2\,\varphi^2\right)
	- \frac {m_{\sst\rm PM}^2} 2\,(\varphi_{\mu\nu}\, \varphi^{\mu\nu} - \varphi^2) +
	\mathscr L^{\sst (2)}_{\sst\rm m.r. }(\varphi,\nabla\varphi)\bigg]\,,
	\label{L PM}
\ea
where the mass of the PM field is given by $m_{\rm\sst PM}^2=\frac 2 3\, \L$\,,
and $\mathscr L^{\sst (2)}_{\sst\rm m.r.}$ is proportional to $G^{\m\n}_{\L}$\,:
\be
	\mathscr L^{\sst (2)}_{\sst\rm m.r.}(\varphi,\nabla\varphi)=
	G_{\L}^{\mu\nu}(a\,\varphi_{\m\r}\,\varphi^{\r}{}_{\nu}+b\,g_{\mu\nu}\,\varphi^{\r\s}\,\varphi_{\r\s}
	+c\,\varphi_{\mu\nu}\,\varphi+d\,g_{\mu\nu}\,\varphi^{2})\,,
\ee
hence arbitrary for the moment. Different choices of $\mathscr L^{\sst (2)}_{\sst\rm m.r.}$
are all physically equivalent as they are related by a field redefinition:
\be
	g_{\m\n} \to g_{\m\n}+\left(\tilde a\,\varphi_{\m\r}\,\varphi^{\r}{}_{\nu}+\tilde b\,g_{\mu\nu}\,\varphi^{\r\s}\,\varphi_{\r\s}
	+\tilde c\,\varphi_{\mu\nu}\,\varphi+\tilde d\,g_{\mu\nu}\,\varphi^{2}\right).
\ee
In eq. \eqref{L PM}, we have also introduced a sign factor $\s$ 
in order to keep track of the role of the relative sign between the graviton and PM kinetic terms:
\begin{itemize}
\item
\mt{\s=+1}\,: the kinetic terms of gravity and PM field have a relatively positive sign, hence
the theory may eventually be unitary;
\item 
\mt{\s=-1}\,: the relative sign is negative, so the gravitons and the PM modes are relatively ghost
violating unitarity already at the linearized level.
\end{itemize}
Finally, by plugging the solution \eqref{L PM} into the condition \eqref{Cond1}, we derive the form of $(\delta_{\a}g_{\m\n})^{\sst (1)}$
for each choice of $\mathscr L^{\sst (2)}_{\sst\rm m.r.}$\,.
Without loss of generality,
we choose $\mathscr L^{\sst (2)}_{\sst\rm m.r.}$ with $(a,b,c,d)=(  2 , -\frac 1 2 , -1 , \frac 1 4 )$\,,
to end up with a relatively simple form for $(\delta_{\a} g_{\mu\nu})^{\sst (1)}_{\a}$\,:
\be
\left(\delta_{\a}\, g_{\mu\nu}\right)^{\sst (1)} = 
2\, \s\, \k \left(2\,\nabla_{(\mu}\varphi_{\nu)\rho}
-\nabla_{\rho}\varphi_{\mu\nu}\right) \partial^{\rho}\a\,,
\label{GR var}
\ee
where we use the weight-one normalization convention for (anti-)symmetrization: $T_{(\mu\nu)}=(T_{\m\n}+T_{\n\m})/2$ and $T_{[\mu\nu]}=(T_{\m\n}-T_{\n\m})/2$\,.

To summarize, we have shown that whenever the PM field is interacting with gravity through
the gravitational minimal coupling --- which has been realized by \emph{covariantizing} 
the free PM action ---
the metric tensor transforms under PM gauge transformations as \eqref{GR var}, up to field redefinitions.

\subsection{Cubic part}
\label{sec: cubic}

We turn to the cubic part of the gauge invariance condition \eqref{cond3}:
\be
	\int d^{4}x\sqrt{-g}
	\left( \left(\delta_{\a}\,\varphi_{\mu\nu}\right)^{\sst (0)}\left[\frac{\delta S^{\sst (3)}}{\delta \varphi_{\mu\nu}}\right]
	+\left(\delta_{\a}\,\varphi_{\mu\nu}\right)^{\sst (1)}\left[\frac{\delta S^{\sst (2)}}{\delta \varphi_{\mu\nu}}\right]
	 + \left(\delta_{\a} g_{\mu\nu}\right)^{\sst (2)}G^{\mu\nu}_{\L} \right)=0\,.
	 \label{cubic cond}
\ee
In this case, we aim to identify $S^{\sst (3)}$ together with 
$(\delta_{\a}\,\varphi_{\mu\nu})^{\sst (1)}$ and $(\delta_{\a} g_{\mu\nu})^{\sst (2)}$.
Similar to the quadratic part, one can solve the condition \eqref{cubic cond} by 
properly covariantizing the PM cubic self-interaction derived for the dS background \cite{Zinoviev:2006im}. 
Checking its gauge invariance on general backgrounds, we get
\begin{align}
S^{\sst (3)}=\l\,\int d^{4}x\sqrt{-g}\,&\Big[
\L\left(\tfrac{8}{3}\, \varphi_{\mu}{}^{\rho} \varphi^{\mu \nu} \varphi_{\nu \rho} - 2\, \varphi^{\mu}{}_{\mu} \varphi_{\nu \rho} \varphi^{\nu \rho} + \tfrac{1}{3} \,\varphi^{\mu}{}_{\mu} \varphi^{\nu}{}_{\nu} \varphi^{\rho}{}_{\rho}\right)
\nonumber\\
&\vphantom{\Big[]}- 2\, \varphi^{\mu \nu} \nabla_{\mu}\varphi^{\rho \sigma} \nabla_{\nu}\varphi_{\rho \sigma} + 2\, \varphi^{\mu \nu} \nabla_{\mu}\varphi^{\rho}{}_{\rho} \nabla_{\nu}\varphi^{\sigma}{}_{\sigma} - 3 \,\varphi^{\mu \nu} \nabla_{\nu}\varphi^{\sigma}{}_{\sigma} \nabla_{\rho}\varphi_{\mu}{}^{\rho} 
\nonumber\\
&\vphantom{\Big[]}- 3\, \varphi^{\mu \nu} \nabla_{\nu}\varphi_{\mu}{}^{\rho} \nabla_{\rho}\varphi^{\sigma}{}_{\sigma} + 2\, \varphi^{\mu \nu} \nabla_{\rho}\varphi^{\sigma}{}_{\sigma} \nabla^{\rho}\varphi_{\mu \nu} -\hphantom{3} \, \varphi^{\mu}{}_{\mu} \nabla_{\rho}\varphi^{\sigma}{}_{\sigma} \nabla^{\rho}\varphi^{\nu}{}_{\nu}
\nonumber\\
&\vphantom{\Big[]} - 2\, \varphi^{\mu \nu} \nabla^{\rho}\varphi_{\mu \nu} \nabla_{\sigma}\varphi_{\rho}{}^{\sigma} + 2\, \varphi^{\mu}{}_{\mu} \nabla^{\rho}\varphi^{\nu}{}_{\nu} \nabla_{\sigma}\varphi_{\rho}{}^{\sigma} + 6\, \varphi^{\mu \nu} \nabla_{\nu}\varphi_{\rho \sigma} \nabla^{\sigma}\varphi_{\mu}{}^{\rho}
\nonumber\\
&\vphantom{\Big[]} + 2\, \varphi^{\mu \nu} \nabla_{\rho}\varphi_{\nu \sigma} \nabla^{\sigma}\varphi_{\mu}{}^{\rho} - 2\, \varphi^{\mu \nu} \nabla_{\sigma}\varphi_{\nu \rho} \nabla^{\sigma}\varphi_{\mu}{}^{\rho} - 2\, \varphi^{\mu}{}_{\mu} \nabla_{\rho}\varphi_{\nu \sigma} \nabla^{\sigma}\varphi^{\nu \rho}
\nonumber\\
&\vphantom{\Big[]} + \hphantom{3\,}\varphi^{\mu}{}_{\mu} \nabla_{\sigma}\varphi_{\nu \rho} \nabla^{\sigma}\varphi^{\nu \rho}\Big]\,.
\label{cubic interactions}
\end{align}
See \autoref{sec: cubic bis}
for the ambient-space formulation 
of the cubic interactions along the lines of \cite{Joung:2012rv}.
By plugging the solution \eqref{cubic interactions} into the condition \eqref{cubic cond}, 
the gauge transformations
$(\delta_{\a}\varphi_{\mu\nu})^{\sst (1)}$ and $(\delta_{\a}g_{\mu\nu})^{\sst (2)}$ can be determined straightforwardly.
In particular, the expression for $(\delta_{\a}\varphi_{\mu\nu})^{\sst (1)}$ will be important 
for the forthcoming analysis and it is given by\footnote{The expression 
for $(\delta_{\a}g_{\mu\nu})^{\sst (2)}$ 
can be equally determined, though we shall not use it in later analysis.
It takes the following relatively simple form,
\be
	(\delta_{\a}g_{\mu\nu})^{\sst (2)}
	= 8\,\kappa \,\lambda\,(
	\varphi_{\rho}{}^{\sigma} \nabla_{(\mu}\varphi_{\nu) \sigma}  
	-  \varphi_{(\mu}{}^{\sigma} \nabla_{\nu)}\varphi_{\rho \sigma}
	+ \varphi_{(\mu}{}^{\sigma} \nabla_{\rho}\varphi_{\nu) \sigma}
	- \varphi_{\rho}{}^{\sigma}  \nabla_{\sigma}\varphi_{\mu \nu})\,\partial^{\rho}\alpha\,,
	\label{metric quadratic}
\ee
after the redefinition,
\ba
g_{\mu\nu} &\to & g_{\mu\nu}+\kappa \lambda\,(12 \, \varphi_{\mu}{}^{\rho} \varphi_{\nu \rho} \varphi^{\sigma}{}_{\sigma}-16 \, \varphi_{\mu}{}^{\rho} \varphi_{\nu}{}^{\sigma} \varphi_{\rho \sigma} + 4\,  \varphi_{\mu \nu} \varphi_{\rho \sigma} \varphi^{\rho \sigma} + \tfrac{20}{3}  g_{\mu \nu} \varphi_{\rho}{}^{\a} \varphi^{\rho \sigma} \varphi_{\sigma \a} 
\nonumber\\
&& \hspace{50pt} -\,4\,  \varphi_{\mu \nu} \varphi^{\rho}{}_{\rho} \varphi^{\sigma}{}_{\sigma} - 6 \, g_{\mu \nu} \varphi^{\rho}{}_{\rho} \varphi_{\sigma \a} \varphi^{\sigma \a} + \tfrac{4}{3} \, g_{\mu \nu} \varphi^{\rho}{}_{\rho} \varphi^{\sigma}{}_{\sigma} \varphi^{\a}{}_{\a})\,.
\ea
}
\be
	(\delta_{\a}\varphi_{\mu\nu})^{\sst (1)}
	= 2\,\s\,\l\left(\nabla_{(\m}\varphi_{\n)\r}-\nabla_{\r}\varphi_{\m\n}\right)
	\partial^{\r}\a\,.\label{PM linear}
\ee
Let us remind the reader that the expression \eqref{cubic interactions} for $S^{\sst (3)}$  is the covariantization of 
the unique two-derivative self-interaction which exists only in four dimensions.
On the other hand, higher-derivative PM self-interactions 
are shown  \cite{Joung:2012rv,Joung:2012hz} to not
affect the form of $(\delta_{\a}\varphi_{\mu\nu})^{\sst (1)}$.
Therefore, the expression \eqref{PM linear} provides the only possible form for the $\varphi$-linear part of nonlinear PM gauge transformation, up to redefinitions of $\varphi_{\m\n}$ which are physically irrelevant.

Notice that the cubic-order gauge-invariance condition \eqref{cond3}
does not constrain the coupling constant $\l$ at all.
The coupling constants can be determined by the quartic or higher-order consistency conditions.
Hence, in principle, we may have to proceed to higher orders to see the eventual
(in-)consistency of the PM-plus-gravity theory.
However, 
there exist other consequences of gauge invariance 
that cubic couplings must satisfy.
They can be examined without analyzing quartic interactions.
In the following, we shall explain this point and solve the correponding conditions.

\section{Symmetries of the PM plus gravity theory}
\label{Sec: symmetries}

Until now, we have analyzed the gauge invariance of the PM plus gravity action up to the cubic order in the PM field.
In general, when an action $S$\,, involving a set of bosonic fields $\chi_{i}$\,, admits gauge symmetries, then 
the gauge symmetries must form an (open) algebra:
\be
	\delta_{\varepsilon}\,\delta_{\eta}
	-\delta_{\varepsilon}\,\delta_{\eta}=
	\delta_{[\eta,\varepsilon]}+(\rm{trivial})\,,
	\label{general}
\ee
where $\d_{\varepsilon}$ stands for \mt{\d_{\varepsilon}=\d_{\varepsilon}\chi_{i}\,\frac{\d\ }{\d \chi_{i}}}
in deWitt notation.
The gauge-algebra bracket $[\eta,\varepsilon]$ might in principle also depend on fields:
$[\eta,\varepsilon]=f(\eta,\varepsilon,\chi_{i})$\,,
while the term ``(trivial)'' denotes any trivial symmetry generated by
an arbitrary antisymmetric matrix $C_{ij}=-C_{ji}$ as
\be
	({\rm trivial})=C_{ij}(\eta,\varepsilon)\,\frac{\d S}{\d\chi_{i}}\,\frac{\d\ }{\d\chi_{j}}\,.
\ee
In the following, we will seek the consequences of the above condition 
for the PM plus gravity theory.

\subsection{Algebra of gauge symmetries}

In the previous sections, we have identified the 
PM gauge transformations up to linear order in the PM fields:
see \eqref{PM zero}, \eqref{GR var} and \eqref{PM linear}.
This makes it possible to identify the $\varphi_{\m\n}$-independent part of the brackets 
by explicitly evaluating the commutator of two successive gauge transformations.
First, the diffeomorphisms give rise to the usual Lie derivative:
\be
	[\,\xi_{2}\,, \xi_{1}\,]=
	\left(\,\xi_{2}^{\nu}\,\nabla_{\nu}\xi_{1}^{\mu}
	- \xi_{1}^{\nu}\,\nabla_{\nu}\xi_{2}^{\mu}\,\right)\partial_{\mu}\,.
	\label{diff diff}
\ee
Next, the commutators between diffeomorphism and PM transformations give
\ba
	\left(\delta_{\a}\,\delta_{\xi}-\delta_{\xi}\,\delta_{\a}\right) 
	g_{\mu\nu}\eq \mathcal O(\varphi)\,,\nn
	\left(\delta_{\a}\,\delta_{\xi}-\delta_{\xi}\,\delta_{\a}\right) 
	\varphi_{\mu\nu} \eq
	 \left(\nabla_{\mu}\nabla_{\nu}+\frac\L3\,g_{\mu\nu}\right) (\xi^\s \,\partial_\s \a)+\mathcal O(\varphi).
\ea
From the above, one can extract the corresponding bracket as
\be
	[\,\x\,, \a\,]=\xi^{\mu}\,\partial_{\mu}\a+\mathcal{O}(\varphi)\,.
	\label{diff PM}
\ee
Finally, there is the commutator of two PM transformations:
its action on $g_{\m\n}$ is given by
\ba
	\left(\delta_{\a_{1}}\,\delta_{\a_{2}}
	-\delta_{\a_{2}}\,\delta_{\a_{1}}\right) g_{\mu\nu} \eq  - \s\,\k 
	\Big[\nabla_\m\left(\partial_\r \a_{2}\,\nabla_{\nu}\partial^\r\a_{1}
	-\partial_\r\a_{1}\,\nabla_{\nu}\partial^\r\a_{2}\right)\nn
	&& \qquad\quad+\, \nabla_\n \left(\partial_\r \a_{2}\,\nabla_{\mu}\partial^\r\a_{1}
	-\partial_\r\a_{1}\,\nabla_{\mu}\partial^\r\a_{2}\right)\Big]+\mathcal{O}(\varphi)\,.
	\label{aa g}
\ea
For the action on $\varphi_{\m\n}$\,,
let us notice that the transformation \eqref{PM linear} involves the covariantization of the tensor,
\be
	C_{\m\n,\r}=\bar\nabla_{(\m}\varphi_{\n)\r}-\bar\nabla_{\r}\varphi_{\m\n}\,,
\ee
which is invariant under the free PM symmetry \eqref{PM sym lin}.
Indeed, one can show that
\be
	\left(\delta_{\a_{1}}\,\delta_{\a_{2}}
	-\delta_{\a_{2}}\,\delta_{\a_{1}}\right) \varphi_{\mu\nu} 
	= \mathcal{O}(\varphi)\,.
	\label{aa phi}
\ee
The absence of $\varphi_{\m\n}$-independent part in the above commutator implies that the bracket 
between two PM transformations does not give a PM transformation, while
 eq.~\eqref{aa g} shows that it 
 results in a diffeomorphism:
\be
	[\,\a_{2}\,, \a_{1}\,]= - \s\,\k \left(\partial_\r \a_{2}\,\nabla^{\mu}\partial^\r\a_{1}
	-\partial_\r\a_{1}\,\nabla^{\mu}\partial^\r\a_{2}\right)\partial_{\mu}+\mathcal O(\varphi)\,.
	\label{PM PM}
\ee
So far, we have determined the $\varphi_{\m\n}$-independent 
part of the gauge-algebra brackets. 
Due to the (possible) field-dependent pieces, the full gauge-algebra brackets do not define a Lie algebra.
However, their restriction to the Killing fields, namely the global-symmetry brackets, \emph{must} 
define a Lie algebra. We will discuss this point in more detail in the next section.

\subsection{Lie algebra of global symmetries}
\label{sec: global symmetries}

Once we get the field-independent part of the gauge-algebra brackets,
it is already sufficient to fully determine the 
global-symmetry structure constants.
Similarly to the gauge symmetries, 
the global-symmetry transformations
must be closed;
what is more is that they must also form a Lie algebra.
Hence, this point --- whether the brackets 
indeed satisfy the Jacobi identity and define a Lie algebra --- 
provides us with a simple necessary condition for 
the consistency of the theory. 

In order to see this point more clearly, let us briefly move back to the 
general discussions presented at the beginning of \autoref{Sec: symmetries}.
We shall now analyze the closure of the symmetry algebra perturbatively.
One considers the expansions:
\ba
	&S=S^{\sst [2]}+S^{\sst [3]}+\cdots\,,\qquad
	&\delta_{\varepsilon}=\d^{\sst [0]}_{\varepsilon}+\d^{\sst [1]}_{\varepsilon}+\cdots\,,\nn
	& [\eta,\varepsilon]=[\eta,\varepsilon]^{\sst [0]}+[\eta,\varepsilon]^{\sst [1]}+\cdots\,,
	\qquad
	& C_{ij}=C_{ij}^{\sst [0]}+C_{ij}^{\sst [1]}+\cdots\,,
\ea
where the superscript $[n]$ stands for the total power of fields $\chi_{i}$\, involved.
Then, the lowest-order part of the closure condition \eqref{general} reads simply
\be
	\delta^{\sst[0]}_{\varepsilon}\,\delta^{\sst [1]}_{\eta}
	-\delta^{\sst [0]}_{\eta}\,\delta^{\sst [1]}_{\varepsilon}=
	\delta^{\sst [0]}_{[\eta,\varepsilon]^{\sst [0]}}\,.
	\label{zeroth closure}
\ee
At the next-to-lowest order, it gives
\be
	 \delta^{\sst[1]}_{\varepsilon}\,\delta^{\sst [1]}_{\eta}
	-\delta^{\sst [1]}_{\eta}\,\delta^{\sst [1]}_{\varepsilon}
	+\delta^{\sst[0]}_{\varepsilon}\,\delta^{\sst [2]}_{\eta}
	-\delta^{\sst [0]}_{\eta}\,\delta^{\sst [2]}_{\varepsilon}
	=\delta^{\sst [1]}_{[\eta,\varepsilon]^{\sst [0]}}
	+\d^{\sst [0]}_{[\eta,\varepsilon]^{\sst [1]}}
	+C^{\sst [0]}_{ij}(\eta,\varepsilon)\,\frac{\d S^{\sst [2]}}{\d\chi_{i}}\,\frac{\d\ }{\d\chi_{j}}\,.
	\label{linear closure}
\ee
Restricting gauge parameters to Killing fields, the 
above two conditions \eqref{zeroth closure} and \eqref{linear closure} provide 
simple but important consistency requirements for the theory.
The Killing fields $\bar\varepsilon$ are defined by the solutions of the Killing equations:
\be
	\d^{\sst [0]}_{\bar\varepsilon}=0\,. 
\ee
The first condition \eqref{zeroth closure} becomes
\be
	\delta^{\sst [0]}_{[\bar\eta,\bar\varepsilon]^{\sst [0]}}=0\,,
\ee
meaning that the global symmetry is closed under the bracket 
$[\![\bar\eta,\bar\varepsilon]\!]:=[\bar\eta,\bar\varepsilon]^{\sst [0]}$\,.
The second condition \eqref{linear closure} reduces to
\be
	 \delta^{\sst[1]}_{\bar\varepsilon}\,\delta^{\sst [1]}_{\bar\eta}
	-\delta^{\sst [1]}_{\bar\eta}\,\delta^{\sst [1]}_{\bar\varepsilon}
	=\delta^{\sst [1]}_{[\![\bar\eta,\bar\varepsilon]\!]}
	+\d^{\sst [0]}_{[\bar\eta,\bar\varepsilon]^{\sst [1]}}
	+C^{\sst [0]}_{ij}(\bar\eta,\bar\varepsilon)\,\frac{\d S^{\sst [2]}}{\d\chi_{i}}\,\frac{\d\ }{\d\chi_{j}}\,,
	\label{admiss}
\ee
meaning that $ \delta^{\sst[1]}_{\bar\varepsilon}$ provides a  representation
of the Lie algebra of the global symmetries on the space of fields.

Having the above general lessons in mind, let us come back to the PM-plus-gravity theory
and consider the dS metric \mt{g_{\m\n}=\bar g_{\mu\nu}}
and \mt{\varphi_{\mu\nu}=0} as the background.
The global symmetries of this background are the subset of gauge symmetries 
which leave it invariant.
The gauge parameters of the global transformations
are defined as the solutions of the following Killing equations:
\be
	\big[\,\delta_{\bar \xi}\,g_{\mu\nu}\,\big]_{\rm\sst bg}=
	2\,\bar\nabla_{(\mu}\bar \xi_{\nu)}=0\,,
	\qquad
	\big[\,\delta_{\bar \a}\,\varphi_{\mu\nu}\,\big]_{\rm\sst bg}
	=\left(\bar\nabla_{\mu}\bar\nabla_{\nu}+
	\frac\L3\,\bar g_{\mu\nu}\right)\bar\a=0\,,
	\label{Killing}
\ee
where $[\,\cdot\,]_{\rm\sst bg}$ means the evaluation $g_{\mu\nu}=\bar g_{\mu\nu}$ and $\varphi_{\mu\nu}=0$\,.
From the lowest part of the gauge-algebra brackets \eqref{diff diff}, \eqref{diff PM} and \eqref{PM PM},
we get the following brackets of global symmetries:
\be
	 [\![\,\bar\varepsilon_{2}\,,\,\bar\varepsilon_{1}\,]\!] =
	2\left(\bar\xi_{[2}^{\nu}\,\partial_{\nu}\,\bar\xi_{1]}^{\mu}\,
	- \s\,\k\,\frac{\L}3\, \bar\a_{[2}\,\partial^\m\bar\a_{1]}\right)
	\partial_{\mu}+2\,\bar\xi_{[2}^{\mu}\,\partial_{\mu}\bar\a_{1]}\,,
	\label{global bracket}
\ee
where we have conveniently packed the parameters as  
$\bar\varepsilon=\bar\xi^{\mu}\,\partial_{\mu}+\bar\a$\,.
The $\bar\xi^{\m}$-transformations form the isometry algebra of dS space,
while the $\bar\a$-transformation extends the isometry to a larger global symmetry.
In order to identify such global symmetry, we need to solve the Killing equations \eqref{Killing}.
For that, it is convenient to reformulate them 
in the ambient-space formalism through the standard embedding:
\be
	\xi_{\mu}(x)=\ell^2\,\frac{\Xi_{\sst M}(X)\,\partial_{\mu}X^{\sst M}}{X^{2}}\,,\qquad
	\a(x)=\ell\,\frac{A(X)}{\sqrt{X^{2}}}\,,
	\label{Ambient gauge}
\ee 
where $\ell^{2}=3/\L$
and the $X^{\sst M}$'s are the coordinates of the ambient space containing dS space as a hyperboloid:
\be
	dS_{4}=\big\{\,X\in \mathbb R^{1,4}\,\big|\,X^{2}=\ell^{2}\,\big\}\,,
	\qquad
	dX^{2}=dR^{2}+\frac{R^{2}}{\ell^{2}}\,\bar g_{\m\n}\,dx^{\m}\,dx^{\n}\,.
	\label{Ambient space}
\ee 
In terms of the ambient space fields, the Killing equations simply read
\be
	\partial_{(\sst M} \bar\Xi_{{\sst N})}=0\,,\qquad \partial_{\sst M}\,\partial_{\sst N}\,\bar A=0\,,
\ee
and the solutions are given by
\be
	\bar\Xi^{\sst M}\,\partial_{\sst M}=W_{\sst AB}\,M^{\sst AB}\,,
	\qquad \bar A=V_{\sst A}\,K^{\sst A}\,.
	\label{global para}
\ee
Here $W_{\sst AB}=-W_{\sst BA}$ and $V_{\sst A}$ are arbitrary parameters
while $M^{\sst AB}$ and $K^{\sst A}$ are the global symmetry generators:
\be
	M^{\sst AB}=2\,X^{[\sst A}\,\partial^{{\sst B}]}\,,
	\qquad K^{\sst A}=X^{\sst A}\,.
\ee
To recapitulate, the global symmetries are generated by the Killing fields:
\be
	\bar\xi^{\mu}=
	W_{\sst AB}\left(2\,\ell^2\,\frac{X^{\sst [A}\,\partial^{\mu} X^{\sst B]}}{X^{2}}\right),
	\qquad
	\bar \a=V_{\sst A}\left(\ell\,\frac{X^{\sst A}}{\sqrt{X^{2}}}\right).
	\label{sol Killing}
\ee
Using these explicit form of the generators and eq.~\eqref{global bracket}, 
one can calculate their brackets and get
\ba
	\big[\hspace{-3pt}\big[\,M^{\sst AB}\,,\,M^{\sst CD}\,\big]\hspace{-3pt}\big]
	\eq \eta^{\sst AD} M^{\sst BC}+\eta^{\sst BC} M^{\sst AD} - \eta^{\sst AC}\,M^{\sst BD} - \eta^{\sst BD} M^{\sst AC} \,,\nn
	\big[\hspace{-3pt}\big[\,M^{\sst AB}\,,\,K^{\sst C}\,\big]\hspace{-3pt}\big] \eq 
	\eta^{\sst BC}\,K^{\sst A} -\eta^{\sst AC}\,K^{\sst B}\,,\nn 
	 \big[\hspace{-3pt}\big[\,K^{\sst A}\,,\,K^{\sst B}\,\big]\hspace{-3pt}\big] \eq - \frac{\L}{3}\,\s\,\k\,M^{\sst AB}\,.
	 \label{brackets}
\ea
The structure constants do not involve the PM self-interaction coupling constant $\lambda$
as it was manifest already from eq.~\eqref{global bracket}. This means that the PM cubic self-interaction
is Abelian.\footnote{In fact, the brackets \eqref{brackets} also encode the properties of the other cubic interactions. 
First, the gravitational self-interaction is associated with the bracket $[\![ M,M]\!]=M$\,:
the fact that it does not vanish implies that the interaction is non-Abelian.
Second, the gravitational minimal-coupling of PM field is associated 
with the brackets $[\![ M,K]\!]=K$ and $[\![ K,K]\!]=M$\,: it is also a non-Abelian interaction.
Finally, the absence of $[\![ K,K]\!]=K$ structure means that the PM self-interaction is Abelian.}
One can easily check that the above brackets \eqref{brackets} define a simple Lie algebra
for any value of the relative sign $\s$ between the kinetic terms:
\begin{itemize}
\item
for \mt{\s=+1}\,, they define $\mathfrak{so}(1,5)$\,,
\item
for \mt{\s=-1}\,, they define $\mathfrak{so}(2,4)$\,. 
\end{itemize}
These algebras contain the isometry algebra $\mathfrak{so}(1,4)$ generated by $M^{\sst AB}$ as a subalgebra.
Hence, we conclude that \emph{any unitary theory of gravitationally interacting PM fields, if such a theory exists, 
must have the global symmetry $\mathfrak{so}(1,5)$\,.}

\subsection{Admissibility condition}
\label{Sec: Representations}

We are now at the point to examine the condition \eqref{admiss}, 
namely the \emph{admissibility condition}, which implies that
the linearized theory must carry a representation of the global symmetry.
The admissibility condition plays an important role in higher-spin field theories \cite{Konstein:1989ij}
as well as in supergravities. 
In the case of PM plus gravity, it will also turn out to be a decisive condition. 

In order to examine the admissibility condition for the system under consideration,
one first needs to linearize the transformations with respect to the metric perturbation 
$h_{\m\n}=g_{\m\n}-\bar g_{\m\n}$ as
\ba
	&& \delta_{\bar\varepsilon}\,h_{\m\n}=\delta_{\bar\varepsilon}\,g_{\m\n}=
	\delta^{\sst [1]}_{\bar\varepsilon}\,h+\mathcal{O}(h,\varphi)\,, \nn
	&& \delta_{\bar\varepsilon}\,\varphi_{\m\n}=\delta^{\sst [1]}_{\bar\varepsilon}\,\varphi
	+\mathcal{O}(h,\varphi)\,, 
\ea
where the superscript $[1]$ means that the corresponding terms are linear in $h_{\m\n}$ or $\varphi_{\m\n}$\,.
First, from the diffeomorphism symmetry, we get 
\be
	\delta^{\sst [1]}_{\bar\xi}\,h_{\m\n} = 2\,\bar\nabla_{(\m}\bar\xi^{\r}\,h_{\n)\r}
	+\bar\xi^{\r}\,\bar\nabla_{\r}h_{\m\n}\,,
	\qquad
	\delta^{\sst [1]}_{\bar\xi}\,\varphi_{\m\n}
	=2\,\bar\nabla_{(\m}\bar\xi^{\r}\,\varphi_{\n)\r}+\bar\xi^{\r}\,\bar\nabla_{\r}\varphi_{\m\n}\,,
\ee
which tell how the $\mathfrak{so}(1,4)$ charges act on the fields.
Then, from eq.~\eqref{PM zero}, \eqref{GR var} and \eqref{PM linear}\,, we get the PM transformations as
\ba
	\delta^{\sst [1]}_{\bar\a}\,h_{\m\n} \eq 
	-2\, \s\, \k\, \partial^{\rho}\bar\a \left(2\,\bar\nabla_{(\mu}\varphi_{\nu)\rho}
	-\bar\nabla_{\rho}\varphi_{\mu\nu}\right),\label{h transf}\\
	\delta^{\sst [1]}_{\bar\a}\,\varphi_{\m\n}
	\eq 2\,\l\,\s\,\partial^{\rho}\bar\a\,\left(\bar\nabla_{(\m}\varphi_{\n)\r}
	-\bar\nabla_{\r}\varphi_{\m\n}\right)-\frac12\,\partial^{\rho}\bar\a \left(2\,\bar\nabla_{(\mu}h_{\nu)\rho}
	-\bar\nabla_{\rho}h_{\mu\nu}\right)+ \frac\Lambda3\,\bar \a\, h_{\m\n}\,.\qquad
	\label{phi transf}
\ea
This indicates how the PM charges $K^{\sst A}$ act on the fields. 
For more explicit expressions, we can replace $\bar\xi^{\m}$ and $\bar\a$ with the 
solutions \eqref{sol Killing} of the Killing equations.
Such expressions can be found in \autoref{sec: rep} where
we carry out the derivation in the ambient-space formalism.

With \eqref{h transf} and \eqref{phi transf}, we are ready to compute the LHS of eq.~\eqref{admiss},
which is the commutator between two PM transformations.\footnote{One can examine also 
the other commutators, but they do actually satisfy the admissibility condition \eqref{admiss}
without constraining 
any coupling constant.} 
After straightforward calculations and imposing the global symmetry condition on gauge parameters, 
we obtain the commutator of two PM transformations as
\ba
	\left(\delta^{\sst [1]}_{\bar \a_{1}}\,\delta^{\sst [1]}_{\bar \a_{2}}
	-\delta^{\sst [1]}_{\bar \a_{2}}\,\delta^{\sst[1]}_{\bar \a_{1}}\right)h_{\m\n}
	\eq 2\,\bar\nabla_{(\m}\mathcal A^{\r}\,h_{\n)\r}+\mathcal A^{\r}\,\bar\nabla_{\r}h_{\m\n}
	+2\,\bar\nabla_{(\m}\mathcal{B}_{\n)}\,, \nn
\left(\delta^{\sst [1]}_{\bar \a_{1}}\,\delta^{\sst [1]}_{\bar \a_{2}}
	-\delta^{\sst [1]}_{\bar \a_{2}}\,\delta^{\sst[1]}_{\bar \a_{1}}\right)\varphi_{\m\n}
	\eq
	\,2\,\bar\nabla_{(\m}\mathcal A^{\r}\,\varphi_{\n)\r}+\mathcal A^{\r}\,\bar\nabla_{\r}\varphi_{\m\n}
	+(\l^2+\s\k)\,\mathcal C_{\m\n}\,,
	\label{CommPhi}
\ea
where $\mathcal A_{\m}$\,, $\mathcal B_{\m}$ and $\mathcal C_{\m\n}$ are given by
\ba
	\mathcal A_\m  \eq 2\,\s\,\k\,\frac{\L}{3}\,\bar\a_{[1}\,\partial_\m\bar\a_{2]}\,,\label{A m}\\
	\mathcal{B}_\m \eq -2\,\s\,\k\left[\partial^{\r}\,\bar\alpha_{[1}\,\partial^{\s}\bar\alpha_{2]}\,
	(\bar\nabla_{\r}h_{\s\m}-4\,\s\,\l\,\bar\nabla_{\r}\varphi_{\s\m})
	+\frac{2\,\L}{3}\, \bar\alpha_{[1}\, \partial^{\rho}\bar\alpha_{2]}\,h_{\rho\m}\right],\\
	\mathcal C_{\m\n}\eq 
	4\,\partial^\rho\alpha_{[1}\,\partial^\s\alpha_{2]}\,\bar\nabla_{(\m|}\bar\nabla_{\s}\varphi_{|\nu)\rho}
	+4\,\L\,\a_{[1}\partial^\r\a_{2]}(\bar\nabla_{(\m}\varphi_{\n)\r}-\bar\nabla_\r\varphi_{\m\n})\,.
\ea
Let us analyze each term in \eqref{CommPhi} 
to see whether they are compatible with the RHS of eq.~\eqref{admiss}:
\begin{itemize}
\item
First, the terms involving $\mathcal A_{\m}$ in \eqref{CommPhi}
take the form of a Lie derivative. 
Moreover one can show that the form \eqref{A m} of $\mathcal A_{\m}$ 
coincides with the bracket \eqref{global bracket}:
\be
	\mathcal A^{\m}\,\partial_{\m}=[\![\, \bar\a_{2}\,,\bar\a_{1}\,]\!]\,.
\ee
Hence, these terms correspond to 
the $\delta^{\sst [1]}_{[\![\,\bar\a_{2}\,\bar\a_{1} ]\!]}$ contribution 
in the RHS of eq.~\eqref{admiss}.
\item
Second, the terms involving $\mathcal B_{\m}$ in \eqref{CommPhi}
take the form of linearized diffeomorphism,
hence corresponding to the 
 $\delta^{\sst [0]}_{[\,\bar\a_{2}\,\bar\a_{1} ]^{\sst [1]}}$ contribution 
in the RHS of eq.~\eqref{admiss} with
\be
	[\, \bar\a_{2}\,,\bar\a_{1}\,]^{\sst [1]}=\mathcal B^{\m}\,\partial_{\m}\,.
\ee
The above relation can be explicitly checked by extracting $\delta^{\sst [2]}_{\a}h_{\m\n}$
from eq.~\eqref{metric quadratic}.

\item
Finally, there remains the $\mathcal C_{\m\n}$ term in \eqref{CommPhi},
which does not correspond to any of the contributions in the RHS of eq.~\eqref{admiss}.
Therefore, in order the admissibility condition to be satisfied, we must require that
the coefficient of the $\mathcal C_{\m\n}$ term vanishes:
\be
	\l^{2}+\s\,\k=0\,.
\ee
This determines the coupling constant $\l$ for the PM self-interaction in terms of the gravitational constant
\mt{\k=8\,\pi\,G_{N}} as \mt{\l=\pm \sqrt{-\s\,\k}}\,.
Now one has two options for a theory of PM-plus-gravity theory depending on the relative sign $\s$ between the kinetic terms: 
\begin{itemize}
\item
for \mt{\s=-1}, we get $\l=\pm\sqrt{\k}$ which coincides with the coupling constant of the PM self-interaction in CG;
\item
for \mt{\s=+1}, the coupling constant $\l$ becomes purely imaginary.
This choice simply corresponds to the $\varphi_{\m\n}\to \pm\,i\,\varphi_{\m\n}$ redefinition 
of CG.
\end{itemize}
\end{itemize}
In this section, we have shown how the admissibility condition allows us to determine 
the Abelian interaction --- the cubic self-interaction of PM field. 
In particular, this demonstrates that the relatively positive kinetic terms in PM plus gravity theory 
cannot be compatible with a real Lagrangian.

\section{Conclusions}
\label{sec: Conclusions}

In this paper, we have investigated the most general form of  Lagrangian
for PM field and gravity with a positive cosmological constant.
By examining its gauge symmetries, we have shown that
\begin{itemize}
\item
There cannot exist a real-valued Lagrangian whose kinetic terms have the
relatively positive sign for 
the PM field and graviton.
In other words, there cannot exist a unitary theory for PM plus gravity.
\item
The PM cubic self-interaction is entirely fixed by gauge invariance. For the relatively negative sign of the kinetic terms, the linearization of this theory admits the global symmetry of $\mathfrak{so}(2,4)$,
which is the conformal algebra in four dimensions.
\item
The case of PM theory without gravity is covered by taking the limit $\k \rightarrow 0$ and choosing the background to be dS space. This limit effectively freezes out the dynamics of the metric tensor. In such a case, the global symmetry reduces to an Abelian one instead of $\mathfrak{so}(1,5)$. The admissibility condition then requires the PM cubic coupling constant $\lambda$ to vanish; therefore, no two-derivative cubic interaction is consistent in the pure PM theory. This rules out the PM limit of massive gravity from the possible consistent theories of the PM field due to the presence of its two- derivative cubic interaction inherited from the Einstein- Hilbert term.
\end{itemize}
Our results imply in particular that the PM limit of the bimetric gravity cannot have
the putative gauge symmetries of PM field. 
Besides, for the relative negative sign of kinetic terms, CG has been recovered as the result of the analysis up to the cubic order in $\varphi_{\m\n}$\,. 
However, we argue that the conclusion can be valid beyond the cubic order, without any assumption on the number of derivatives for the following reasons:
\begin{itemize}
\item
The inclusion of higher-derivative interactions cannot change the conclusion
since they do not affect the form of PM transformations \cite{Joung:2012rv}, on which our analysis is based on.
\item
The validity beyond the cubic order can be argued from the two points:
\emph{first,} CG is the unique Weyl invariant as well as 
the unique invariant theory of $\mathfrak{so}(2,4)$-gauge connection \cite{Kaku:1977pa};
\emph{second,} the interaction structure of CG is the unique one up to four derivatives \cite{Boulanger:2001he}.
\end{itemize}
We expect that a similar construction might be possible in higher dimensions 
where the corresponding CG equations still have a factorized form on any Einstein background 
\cite{Nutma:2014pua} involving some massive modes.

\acknowledgments{
EJ thanks J. Mourad for useful discussions. MT thanks Stefan Theisen for useful discussions. MT is also grateful to KITPC for the hospitality and support during the final stages of this work.}

\appendix

\section{PM bimetric gravity as a theory of gravity plus matter}
\label{sec: bimetric}
In this section, we show how the bimetric gravity of \cite{Hassan:2011ea,Hassan:2011zd}
can be recast into the form  \eqref{Action}.
For simplicity, let us consider the model with a particular choice of parameters,
which was considered as the unique candidate for the PM plus gravity theory in that context.
Its action takes the following form after a rescaling of the metric $f_{\mu\nu}\rightarrow \big(\frac {m_g}{m_f}\big)^2 f_{\mu\nu}$\,,
\be
	S[g,f]=\frac{1}{2\,\k}\int d^{4}x\left[
	\sqrt{-g} \left(R(g)-2\L\right)+
	\sqrt{-f} \left(R(f)-2\L\right)
	-\frac{\L}3\,V(g,f)\right],
\ee
where the potential term $V(g,f)$ is given by
\be
	V(g,f)=\sqrt{-g}\left[({\rm Tr} S)^{2}-{\rm Tr} S^{2}\right],
	\qquad
	(S^2)^{\mu}{}_{\nu}=g^{\m\r}\,f_{\r\n}\,.
\ee
The  constants $\k$ and $\L$ are related to the parameters used in \cite{Hassan:2012rq}
as follows,
\be
	2\,\k= \frac1{m_g^{2}}\,,
	\qquad 
	\L=3\,\frac {m^4}{m_f^2} 
	\,\beta_2\,.
\ee
Then one can redefine its two tensors
into the physical metric $G_{\m\n}$
and the massive spin-two field $\varphi_{\m\n}$ as
\be\label{redef}
	g_{\m\n}=\frac12\left(G_{\m\n}+2\,\varphi_{\m\n}\right),
	\qquad
	f_{\m\n}=\frac12\left(
	G_{\m\n}- 2 \,\varphi_{\m\n}\right),
\ee
in order to recover an action $S[G,\varphi]$ of the form \eqref{Action}.
This action can be derived by expanding $S[g,f]$ around the \emph{background} $g_{\m\n}=G_{\m\n}/2=f_{\m\n}$  
and by replacing the \emph{fluctuation} by the PM field $\varphi_{\mu\nu}$\,:
the $\varphi^n$-order part of the action
is given by
\be
	S^{\sst (n)}[G,\varphi]=\frac1{n!}
    \left[\int\varphi_{\m\n}\left(
	\frac {\delta}{\delta g_{\m\n}}
	-\frac {\delta}{\delta f_{\m\n}}\right)
\right]^n S[g,f]\Big|_{g_{\m\n}=\frac12\,G_{\m\n}=f_{\m\n}}\,.
\ee
For example, the zeroth order action $S^{\sst (0)}$
gives the Einstein-Hilbert action:
\be
	S^{\sst (0)}=S[g,f]\Big|_{g_{\m\n}=\frac12\,G_{\m\n}
	=f_{\m\n}}=\frac{1}{2\,\k}\int d^{4}x\,\sqrt{-G}\,\big[R(G)-2\,\L\big].
\ee
At the first order, we get $S^{\sst (1)}=0$ due to the 
symmetry,
\be
    S[g,f]=S[f,g]\,.
    \label{bimetric sym}
\ee
The quadratic action $S^{\sst (2)}$
linearized around dS space was 
already computed in \cite{Hassan:2012rq}, which confirms that it coincides with the PM free Lagrangian. 
Therefore, one can see that the action of the bimetric gravity can be recast into the stardard form of the \emph{gravity plus matter}, which our analysis concerned.
Moreover, one can also see, from the symmetry \eqref{bimetric sym}, that all odd powers of PM field do not appear in the action $S[G,\varphi]$\,.

\section{Cubic-interaction analysis in ambient-space formulation} 

In \autoref{sec: cubic} and   \autoref{Sec: Representations},
we provided the form of the PM cubic self-interaction \eqref{cubic interactions},
and  computed the commutator of the linearized transformations \eqref{CommPhi}.
In this appendix, we present the ambient-space version of such calculations, which makes 
the dS symmetry and the link to the previous works \cite{Joung:2012rv,Joung:2012hz} more transparent.
For that, we first recast everything into the objects in the ambient space \eqref{Ambient space}:
the metric perturbation $h_{\m\n}$ and the PM field $\varphi_{\m\n}$ 
correspond to the ambient-space fields $H(X,U)$ and $\Phi(X,U)$ as
\be
	h_{\m\n}(x)=\frac{\partial_{\m}X^{\sst M}\,\partial_{\n}X^{\sst N}}{X^{2}}\,
	\frac{\partial^{2}H(X,U)}{\partial U^{\sst M}\partial U^{\sst N}}\,,\qquad
	\varphi_{\m\n}(x)=\frac{\partial_{\m}X^{\sst M}\,\partial_{\n}X^{\sst N}}{\sqrt{X^{2}}}\,
	\frac{\partial^{2}\Phi(X,U)}{\partial U^{\sst M}\partial U^{\sst N}}\,,
\ee
while the corresponding gauge parameters are already given in eq.~\eqref{Ambient gauge}.
Here, we have also introduced the auxiliary variables $U^{\sst M}$ to handle the tensor indices conveniently.

\subsection{Cubic interactions of PM field}

\label{sec: cubic bis}

Following \cite{Joung:2012fv} (see \cite{Taronna:2012gb} for more details), the part of the cubic interactions  not involving traces and divergences can be obtained as a function
 of six independent scalar contractions:
\be
 Z_i=\partial_{U_{i+1}}\!\!\cdot\partial_{U_{i-1}}\,,\qquad Y_i=\partial_{U_i}\!\cdot\partial_{X_{i+1}}
 \qquad [i\simeq i+3]\,.
\ee
About the two-derivative cubic interactions considered in this paper, it reads
\ba
	&& \int_{\rm dS}\bigg[\ 
	\frac{1}{2\k}\left[ (Y_1\,Z_1+Y_2\,Z_2+Y_3\,Z_3)^2+3\,\L \,Z_1\, Z_2\, Z_3\right]
	H_{1}\,H_{2}\,H_{3}\nn
	&&\hspace{30pt}+\,
	\frac{\s}{2}\left[(Y_1\,Z_1+Y_2\,Z_2+Y_3\,Z_3)^2+\L\,Z_1\,Z_2\,Z_3\right]
	H_{1}\,\Phi_{2}\,\Phi_{3}\nn
	&&\hspace{30pt}+\,
	\frac{\l}{2}\,(Y_1\,Z_1+Y_2\,Z_2+Y_3\,Z_3)^2\,
	\Phi_{1}\,\Phi_{2}\,\Phi_{3}\,\bigg]_{{}^{X_{i}=X}_{U_{i}=0}}\,,
	\label{Amb Cub}
\ea
where $H_{i}$ and $\Phi_{i}$ stand for $H(X_{i},U_{i})$ and $\Phi(X_{i},U_{i})$\,.
This form of the action is invariant under the gauge transformation, 
\be
	\delta\,H=U\cdot\partial_{X}\,\Xi\,,\qquad
	\delta\,\Phi=(U\cdot\partial_{X})^{2}\,A\,,
\ee
modulo the terms involving divergences and traces. Several comments are in order:
\begin{itemize}
\item
First, the $H_{1}\,H_{2}\,H_{3}$ term corresponds to the usual gravitational self-interactions.
\item
Second, the $H_{1}\,\Phi_{2}\,\Phi_{3}$ term corresponds the the gravitational minimal coupling
of the PM field. In fact, as far as the TT part is concerned \cite{Joung:2012rv}, there exists one more two-derivative coupling of the form:
\be
	(Y_{1}\,Z_{1}+Y_2\,Z_2+Y_3\,Z_3)(Y_{1}\,Z_{1}-Y_2\,Z_2-Y_3\,Z_3)\,.
\ee
However, there do not exist
divergence and trace pieces that can uplift the above coupling to a fully gauge invariant one.

\item
Finally, the $\Phi_{1}\,\Phi_{2}\,\Phi_{3}$ term corresponds to
the PM self interaction, which exists only in four dimensions.
The work \cite{Joung:2012rv} does not contain this coupling since
it only concerns the interactions existing in generic dimensions.
A particular dimensional dependence in the latter work was hidden in the variable $\hat\delta$\,,
which can be eventually replaced as
\be
	\int_{\rm dS} {\hat\delta}^{\,n}\,I_{\Delta}
	=\frac{(\D+d-1)(\D+d-3)\cdots (\D+d-2n+1)}{\ell^{n}}\,\int_{\rm dS}\,I_{\Delta}\,,
\ee
where the integrand $I_{\D}$ satisfies $(X\cdot\partial_{X}-\D)\,I_{\D}=0$\,. 
Taking into account the above identity, one can easily find the two-derivative
$\Phi_{1}\,\Phi_{2}\,\Phi_{3}$ coupling.
Let us also rectify one claim in the literature ---
the footnote 7 of \cite{Deser:2012qg} and the footnote 11 of \cite{Joung:2012rv}:
the four-derivative PM self-interaction does \emph{not} reduce
to the two-derivative one due to the Gauss-Bonnet identity,
but it actually coincides with the Gauss-Bonnet term,
hence vanishing identically in four dimensions.

\end{itemize}

\subsection{Global symmetries}
\label{sec: rep}

After identifying the cubic interactions in the ambient-space form,
one can systematically extract 
the corresponding deformations of gauge transformations following \cite{Joung:2013nma}.
Upon restricting on global symmetries
adopting a \mt{\ell=1} convention, 
we obtain the following set of deformations from the expression \eqref{Amb Cub}
and the results obtained in \cite{Joung:2013nma}:
\ba
	\delta^{\sst [1]} H \eq -\Pi
	\left[\,U\cdot\partial_{U_2} \left(\frac{1}{2}\,U\cdot\partial_{U_2}\,
	\partial_{U_1}\!\cdot\partial_{X_2}-U\cdot\partial_{U_1}\,\partial_{U_2}\!\cdot\partial_{X_1}\right)
	\Xi_1\,H_2\right.\nn
	&& \hspace{25pt} \left.+\, \s\,\k\,(U\cdot\partial_{U_2})^2
	\left(\partial_{X_1}\!\cdot\partial_{X_2}-\frac{1}{2}\right) 
 	A_1\,\Phi_2\,\right], \\
	\delta^{\sst [1]} \Phi\eq -\Pi\left[\,
	U\cdot\partial_{U_2}\left(\frac{1}{2}\,U\cdot\partial_{U_2}\,\partial_{U_1}\!\cdot\partial_{X_2}
	-U\cdot\partial_{U_1}\,\partial_{U_2}\!\cdot\partial_{X_1}\right) \Xi_1\,\Phi_2\right.\nn
	&&\hspace{10pt} \left.+\,\frac{1}{4}\,(U\cdot\partial_{U_2})^2
	\left(\partial_{X_1}\!\cdot\partial_{X_2} -\frac{1}{2}\right) A_1\,H_2
	+\frac{\l}{2}\,(U\cdot\partial_{U_2})^2\,\partial_{X_1}\!\cdot\partial_{X_2}\,A_1\,\Phi_2\,
	\right],
\ea
where $\Pi$ is the operator adjusting tangent or radial contributions
so that  the resulting deformations of the gauge transformations remain compatible with the tangentiality and homogeneity constraints.
Making use of the solution \eqref{global para},
the $M^{\sst AB}$-transformations of the isometry algebra $\mathfrak{so}(1,4)$ with parameters $W_{\sst AB}$ become
\ba
	\delta^{\sst [1]}_{W}\,H
	\eq \left(X\cdot W\cdot \partial_{X}
	+U\cdot W\cdot \partial_{U}\right) H\,, \nn
	\delta^{\sst [1]}_{W}\,\Phi
	\eq \left(X\cdot W\cdot \partial_{X}
	+U\cdot W\cdot \partial_{U}\right) \Phi\,.
\ea
On the other hand, the $K^{\sst A}$-transformations, associated to the 
additional generators of the global symmetries with parameters $V_{\sst A}$, take the form,
\ba
	\delta^{\sst [1]}_{V}\,H
	\eq -2\,\s\,\k\left(X^{2}\right)^{\frac32}
	\left(U\cdot\partial_{X}\,V\cdot\partial_{U}-V\cdot\partial_{X}+2\,\frac{X\cdot U}{X^{2}}\,V\cdot\partial_{U}
	\right)\left(X^{2}\right)^{-\frac12}\,\Phi\,, \\
	\delta^{\sst [1]}_{V}\,\Phi
	\eq -\frac12 \left(U\cdot\partial_{X}\,V\cdot\partial_{U}-V\cdot\partial_{X}\right)H
	+\l\left(X^{2}\right)^{\frac12}
	\left(U\cdot\partial_{X}\,V\cdot\partial_{U}-2\,V\cdot\partial_{X}\right)\Phi\,.\quad
\ea
Using these expressions of the gauge transformations, one can easily compute the relevant commutators.
In particular, we are interested in the commutator between two $K^{\sst A}$-transformations $\delta^{\sst [1]}_{V_{[2}}\d^{\sst [1]}_{V_{1]}}$\,. After a straightforward computation, one gets
\ba
\delta^{\sst [1]}_{V_{[2}}\d^{\sst [1]}_{V_{1]}}H \eq
-\s\,\k\left(V_{[1}\cdot X\,V_{2]}\cdot \partial_{X}
+V_{[1}\cdot U\,V_{2]}\cdot \partial_{U}\right) H +
U\cdot\partial_{X}\,\cal B\,,\\
\delta^{\sst [1]}_{V_{[2}}\d^{\sst [1]}_{V_{1]}}\Phi \eq
(2\,\s\,\k+3\,\l^{2})\left(V_{[1}\cdot X\,V_{2]}\cdot \partial_{X}
+V_{[1}\cdot U\,V_{2]}\cdot \partial_{U}\right) \Phi+
(\l^{2}+\s\,\k)\,\cal C\,,
\ea
with
\ba
	\cal B \eq \s\,\k\left(X^{2}\,V_{[1}\cdot \partial_{X}\,V_{2]}\cdot \partial_{U}
+V_{[1}\cdot X\,V_{2]}\cdot \partial_{U}\right)H
-4\,\s\,\k\,\l\left(X^{2}\right)^{\frac32}\,V_{[1}\cdot \partial_{X}\,V_{2]}\cdot \partial_{U}\,\Phi \,,\nn
	\cal C\eq
	\Big[U\cdot\partial_{X}\,V_{[1}\cdot \partial_{X}\,V_{2]}\cdot\partial_{U}\,X^{2}
-4\,U\cdot\partial_{X}\,V_{[1}\cdot X\,V_{2]}\cdot\partial_{U}
+X\cdot U\,V_{[1}\cdot \partial_{X}\,V_{2]}\cdot \partial_{U}\nn
&&\quad +\,V_{[1}\cdot X\,V_{2]}\cdot \partial_{X}
-V_{[1}\cdot U\,V_{2]}\cdot \partial_{U}\Big]\,\Phi\,.
\ea
Comparing the above expression with the bracket,
\be
	[\![ \,V_{1}\cdot K\,,\,V_{2}\cdot K\,]\!]
	=-\s\,\k\,V_{1}\cdot M\cdot V_{2}\,,
\ee
and imposing closure one can conclude that the $\cal C$ term has to vanish:
\be
	\l^{2}+\s\,\k=0\,,
\ee
while the $\cal B$ term corresponds to the contribution of $\delta^{\sst [0]}_{[\bar \eta,\bar\varepsilon]^{\sst [2]}}$\,.

\bibliographystyle{JHEP}

\begin{thebibliography}{10}

\bibitem{Deser:1983mm}
S.~Deser and R.~I. Nepomechie, {\it {Gauge invariance versus masslessness in de
  sitter space}},  {\em Annals Phys.} {\bf 154} (1984) 396.

\bibitem{Deser:2001us}
S.~Deser and A.~Waldron, {\it {Partial masslessness of higher spins in (A)dS}},
   {\em Nucl.Phys.} {\bf B607} (2001) 577--604,
  [\href{http://xxx.lanl.gov/abs/hep-th/0103198}{{\tt hep-th/0103198}}].

\bibitem{Higuchi:1986py}
A.~Higuchi, {\it {Forbidden mass range for spin-2 field theory in de sitter
  space-time}},  {\em Nucl.Phys.} {\bf B282} (1987) 397.

\bibitem{deRham:2010ik}
C.~de~Rham and G.~Gabadadze, {\it {Generalization of the Fierz-Pauli Action}},
  {\em Phys.Rev.} {\bf D82} (2010) 044020,
  [\href{http://xxx.lanl.gov/abs/1007.0443}{{\tt arXiv:1007.0443}}].

\bibitem{deRham:2010kj}
C.~de~Rham, G.~Gabadadze, and A.~J. Tolley, {\it {Resummation of Massive
  Gravity}},  {\em Phys.Rev.Lett.} {\bf 106} (2011) 231101,
  [\href{http://xxx.lanl.gov/abs/1011.1232}{{\tt arXiv:1011.1232}}].

\bibitem{Hassan:2011vm}
S.~Hassan and R.~A. Rosen, {\it {On Non-Linear Actions for Massive Gravity}},
  {\em JHEP} {\bf 1107} (2011) 009,
  [\href{http://xxx.lanl.gov/abs/1103.6055}{{\tt arXiv:1103.6055}}].

\bibitem{Hassan:2011ea}
S.~Hassan and R.~A. Rosen, {\it {Confirmation of the Secondary Constraint and
  Absence of Ghost in Massive Gravity and Bimetric Gravity}},  {\em JHEP} {\bf
  1204} (2012) 123, [\href{http://xxx.lanl.gov/abs/1111.2070}{{\tt
  arXiv:1111.2070}}].

\bibitem{Hassan:2011zd}
S.~Hassan and R.~A. Rosen, {\it {Bimetric Gravity from Ghost-free Massive
  Gravity}},  {\em JHEP} {\bf 1202} (2012) 126,
  [\href{http://xxx.lanl.gov/abs/1109.3515}{{\tt arXiv:1109.3515}}].

\bibitem{Bergshoeff:2009aq} 
  E.~A.~Bergshoeff, O.~Hohm and P.~K.~Townsend,
  {\it{More on Massive 3D Gravity}}
  {\em Phys.Rev.} {\bf D79} (2009) 124042,
   [\href{http://xxx.lanl.gov/abs/0905.1259}{{\tt arXiv:0905.1259}}].

\bibitem{deRham:2012kf} 
  C.~de Rham and S.~Renaux-Petel,
  {\it{Massive Gravity on de Sitter and Unique Candidate for Partially Massless Gravity}},
  {\em JCAP} {\bf 1301} (2013) 035, 
  [\href{http://xxx.lanl.gov/abs/1206.3482}{{\tt arXiv:1206.3482}}].

\bibitem{Hassan:2012gz}
S.~Hassan, A.~Schmidt-May, and M.~von Strauss, {\it {On Partially Massless
  Bimetric Gravity}},  {\em Physics Letters} {\bf B726,} (2013) 834,
  [\href{http://xxx.lanl.gov/abs/1208.1797}{{\tt arXiv:1208.1797}}].

\bibitem{Hassan:2012rq}
S.~Hassan, A.~Schmidt-May, and M.~von Strauss, {\it {Bimetric theory and
  partial masslessness with Lanczos--Lovelock terms in arbitrary dimensions}},
  {\em Class.Quant.Grav.} {\bf 30} (2013) 184010,
  [\href{http://xxx.lanl.gov/abs/1212.4525}{{\tt arXiv:1212.4525}}].

\bibitem{Hassan:2013pca}
S.~Hassan, A.~Schmidt-May, and M.~von Strauss, {\it {Higher Derivative Gravity
  and Conformal Gravity From Bimetric and Partially Massless Bimetric Theory}},
   \href{http://xxx.lanl.gov/abs/1303.6940}{{\tt arXiv:1303.6940}}.


\bibitem{Deser:2012qx}
S.~Deser and A.~Waldron, {\it {Acausality of Massive Gravity}},  {\em
  Phys.Rev.Lett.} {\bf 110} (2013), no.~11 111101,
  [\href{http://xxx.lanl.gov/abs/1212.5835}{{\tt arXiv:1212.5835}}].

\bibitem{Deser:2013uy}
S.~Deser, M.~Sandora, and A.~Waldron, {\it {Nonlinear Partially Massless from
  Massive Gravity?}},  {\em Phys.Rev.} {\bf D87} (2013) 101501,
  [\href{http://xxx.lanl.gov/abs/1301.5621}{{\tt arXiv:1301.5621}}].

\bibitem{deRham:2013wv}
C.~de~Rham, K.~Hinterbichler, R.~A. Rosen, and A.~J. Tolley, {\it {Evidence for
  and Obstructions to Non-Linear Partially Massless Gravity}},  {\em Phys.Rev.}
  {\bf D88} (2013) 024003, [\href{http://xxx.lanl.gov/abs/1302.0025}{{\tt
  arXiv:1302.0025}}].

\bibitem{Deser:2013gpa}
S.~Deser, M.~Sandora, and A.~Waldron, {\it {No consistent bimetric gravity?}},
  {\em Phys.Rev.} {\bf D88} (2013) 081501,
  [\href{http://xxx.lanl.gov/abs/1306.0647}{{\tt arXiv:1306.0647}}].

\bibitem{Fasiello:2013woa}
  M.~Fasiello and A.~J.~Tolley,
 {\it {Cosmological Stability Bound in Massive Gravity and Bigravity}},
  {\em JCAP} {\bf 1312} (2013) 002,
  [\href{http://xxx.lanl.gov/abs/1308.1647}{{\tt arXiv:1308.1647}}].


\bibitem{Fradkin:1981iu}
E.~Fradkin and A.~A. Tseytlin, {\it {Renormalizable asymptotically free quantum
  theory of gravity}},  {\em Nucl.Phys.} {\bf B201} (1982) 469--491.

\bibitem{Lee:1982cp}
S.~Lee and P.~van Nieuwenhuizen, {\it {Counting of States in Higher Derivative
  Field Theories}},  {\em Phys.Rev.} {\bf D26} (1982) 934.

\bibitem{Riegert:1984hf}
R.~Riegert, {\it {The particle content of linearized Conformal Gravity}},  {\em
  Phys.Lett.} {\bf A105} (1984) 110--112.

\bibitem{Maldacena:2011mk}
J.~Maldacena, {\it {Einstein Gravity from Conformal Gravity}},
  \href{http://xxx.lanl.gov/abs/1105.5632}{{\tt arXiv:1105.5632}}.

\bibitem{Deser:2012qg}
S.~Deser, E.~Joung, and A.~Waldron, {\it {Partial Masslessness and Conformal
  Gravity}},  {\em J.Phys.} {\bf A46} (2013) 214019,
  [\href{http://xxx.lanl.gov/abs/1208.1307}{{\tt arXiv:1208.1307}}].

\bibitem{Zinoviev:2006im}
Y.~Zinoviev, {\it {On massive spin 2 interactions}},  {\em Nucl.Phys.} {\bf
  B770} (2007) 83--106, [\href{http://xxx.lanl.gov/abs/hep-th/0609170}{{\tt
  hep-th/0609170}}].

\bibitem{Zinoviev:2013hac}
Y.~Zinoviev, {\it {All spin-2 cubic vertices with two derivatives}},  {\em
  Nucl.Phys.} {\bf B872} (2013) 21--37,
  [\href{http://xxx.lanl.gov/abs/1302.1983}{{\tt arXiv:1302.1983}}].

\bibitem{Zinoviev:2014zka}
Y.~M. Zinoviev, {\it {Massive spin-2 in the Fradkin-Vasiliev formalism. I.
  Partially massless case}},  \href{http://xxx.lanl.gov/abs/1405.4065}{{\tt
  arXiv:1405.4065}}.

\bibitem{Joung:2012rv}
E.~Joung, L.~Lopez, and M.~Taronna, {\it {On the cubic interactions of massive
  and partially-massless higher spins in (A)dS}},  {\em JHEP} {\bf 1207} (2012)
  041, [\href{http://xxx.lanl.gov/abs/1203.6578}{{\tt arXiv:1203.6578}}].

\bibitem{Joung:2012hz}
E.~Joung, L.~Lopez, and M.~Taronna, {\it {Generating functions of
  (partially-)massless higher-spin cubic interactions}},  {\em JHEP} {\bf 1301}
  (2013) 168, [\href{http://xxx.lanl.gov/abs/1211.5912}{{\tt
  arXiv:1211.5912}}].

\bibitem{Boulware:1973my}
D.~Boulware and S.~Deser, {\it {Can gravitation have a finite range?}},  {\em
  Phys.Rev.} {\bf D6} (1972) 3368--3382.

\bibitem{Konstein:1989ij}
S.~Konstein and M.~A. Vasiliev, {\it {Extended Higher Spin Superalgebras and
  Their Massless Representations}},  {\em Nucl.Phys.} {\bf B331} (1990)
  475--499.


\bibitem{Kaku:1977pa}
M.~Kaku, P.~Townsend, and P.~van Nieuwenhuizen, {\it {Gauge Theory of the
  Conformal and Superconformal Group}},  {\em Phys.Lett.} {\bf B69} (1977)
  304--308.

\bibitem{Boulanger:2001he}
N.~Boulanger and M.~Henneaux, {\it {A Derivation of Weyl gravity}},  {\em
  Annalen Phys.} {\bf 10} (2001) 935--964,
  [\href{http://xxx.lanl.gov/abs/hep-th/0106065}{{\tt hep-th/0106065}}].

\bibitem{Nutma:2014pua}
T.~Nutma and M.~Taronna, {\it {On conformal higher spin wave operators}},
  \href{http://xxx.lanl.gov/abs/1404.7452}{{\tt arXiv:1404.7452}}.

\bibitem{Joung:2012fv}
E.~Joung, L.~Lopez, and M.~Taronna, {\it {Solving the Noether procedure for
  cubic interactions of higher spins in (A)dS}},  {\em J.Phys.} {\bf A46}
  (2012) 214020, [\href{http://xxx.lanl.gov/abs/1207.5520}{{\tt
  arXiv:1207.5520}}].

\bibitem{Taronna:2012gb}
M.~Taronna, {\it {Higher-Spin Interactions: three-point functions and beyond}},
   \href{http://xxx.lanl.gov/abs/1209.5755}{{\tt arXiv:1209.5755}}.

\bibitem{Joung:2013nma}
E.~Joung and M.~Taronna, {\it {Cubic-interaction-induced deformations of
  higher-spin symmetries}},  {\em JHEP} {\bf 1403} (2014) 103,
  [\href{http://xxx.lanl.gov/abs/1311.0242}{{\tt arXiv:1311.0242}}].

\end{thebibliography}

\end{document}